# Light-induced Pairing Instability of Ultrafast Electron Beams with Space Charge Interactions

Hao Geng[1,2], Qiaofei Pan[3,4], Jian Kang[5], Yiming Pan[5,6*]

1. College of Physics, Nanjing University of Aeronautics and Astronautics, Nanjing 211106, China
2. Key Laboratory of Aerospace Information Materials and Physics (NUAA), MIIT, Nanjing 211106, China
3. Institute of Precision Optical Engineering, School of Physics Science and Engineering, Tongji University, Shanghai, 200092, China
4. MOE Key Laboratory of advanced micro-structure materials, Shanghai, 200092, China
5. State Key Laboratory of Quantum Functional Materials, School of Physical Science and Technology, ShanghaiTech University, Shanghai 200031, China
6. Center for Transformative Science, ShanghaiTech University, Shanghai 200031, China

**Abstract**

Ultrafast electron beams are essential for many applications, yet space-charge interactions in high-intensity beams lead to energy dissipation, coherence loss, and pulse broadening. Existing techniques mitigate these effects by using low-flux beams, preserving beam coherence into quantum regime. Here, we propose a novel approach by treating the electrons as a strongly-correlated Fermi gas rather than merely as an ensemble of charged point-like particles. We introduce a photon-induced pairing mechanism that generates a net attractive force between two electrons, thereby forming "flying bound states" analogous to Cooper pairs of conduction electrons in superconductors. Employing the setting of photon-induced near-field electron microscopy (PINEM), we demonstrate that the effective interaction via single-photon exchange among PINEM electrons can suppress the inherent repulsive Coulomb interaction, enabling a paring instability mediated by structured electromagnetic fields at near-resonant velocity matching regimes. Finally, we analyze the dynamics of the free-electron pairs in a bunched beam, underscoring the potential to facilitate a phase-coherent condensate of electrons, which can further enhance beam coherence and multi-particle correlation for high-intensity electrons.

Ultrafast electron beams, generated and controlled by femtosecond laser pulses, have become crucial for advancements in areas such as electron microscopy [1], diffraction [2], particle acceleration, attosecond beam bunching [3,4], diverse radiative schemes [5–8] and cathodoluminescence [9]. Beyond these applications, light-induced quantum control of ultrafast electrons has enabled novel fields, including quantum wavefunction engineering [1], both longitudinal and transverse beam shaping [11–15], free electron quantum optics [16], non-perturbative strong-field physics [17], and multiphoton processes [18,19]. These emerging developments are widely explored in recent studies, emphasizing the growing sophistication and relevance of light-electron interactions in ultrafast quantum regimes. Nevertheless, despite these advances, a critical challenge remains in managing the space-charge repulsion between electrons, which restricts the coherence and brightness of ultrafast electron beams compared to photon beams, due to the strong repulsive forces and the constraints imposed by the Pauli's exclusion principle [20,21]. This repulsion results in spectral distortions, decoherence, and pulse degradation in electron beams [2,22–24]. Enhancing the coherence of multi-particle beams would not only expand their applications but also may lead to a fundamentally new phase of light-induced free-electron matter, capable of undergoing a transition akin to those phase transitions observed in solids.

Usually, electrons in a solid can exhibit high coherence at low temperatures, sometimes even possible to condensate to a highly-correlated phase of matter. For instance, for a conventional superconductor, quasi-free electrons can form Cooper pairing via phonon-mediated net attraction, resulting in phase coherence and macroscopically observable effects such as zero resistance and the Meissner effect [25]. Similarly, reducing the space-charge effect to facilitate pairing of free-electrons in ultrafast beams could lead to a light-induced "free-electron superconducting state", characterized by coherent pair condensates (similar to the picture of Bose-Einstein condensate (BEC) in ultracold atoms [26]), thereby permitting high-brightness beams. For this, an effective attractive interaction is necessary to counteract intrinsic space-charge repulsion, leading to a paring instability where, once a net attractive force emerges between free-electrons, two plane-wave electrons attract each other and form a bound state [27].

In this work, we propose that the tailored electromagnetic fields could serve as bosonic mediators, inducing a net attractive interaction and potentially enabling paring instability within ultrafast electron beams, akin to the formation of Cooper pairs in phonon-mediated superconductors. Exploring this pairing phenomenon in free electrons differs from solid-state systems due to the absence of a lattice [which creates phonons by lattice vibrations] or Fermi surface [which defines the relevant electronic band structure], both of which are believed to be essential for conventional superconductivity. For instance, a theoretical study has proposed that electron interactions mediated by cavity photons [18], results in pairing instability and electron-photon superconductivity [28]. This electron-photon superconductivity differs from the recent experimental demonstration of photo-induced superconductivity [29], as the latter still relies on a conventional phonon-based pairing process whose infrared-active phonons and collective vibrational modes of a molecular solid are coherently controlled by laser pulses. Despite these differences, recent advances [14,30,31] in free-electron quantum optics suggest that engineered light-matter interactions may enable new forms of electron interactions and correlations. Studies show that free electrons in presence of light, although distinct from solid-state electrons, exhibit quasi-free behavior of Floquet steady-states akin to conduction-band electrons in solids that possess Bloch wave features, in which the light-induced time periodicity mimics the spatial periodicity of lattice. By analogy, any mechanism that provides a net attractive interaction between electrons, such as structured electromagnetic vacuum (e.g., cavity QED, nanoplasmonics), could induce pairing behavior in electrons, even in free space, a phenomenon first described by L. Cooper [27]. This perspective gives us confidence to broaden the concept of pairing, spanning processes from electron pairs in solids to photon pairing in nonlinear optics and to atom pairs in ultracold gases.

Our work specifically investigates the pairing instability in ultrafast free electrons modulated through the setting of photon-induced near-field electron microscopy (PINEM) [11]. At an almost resonant and weak-field regime, we find that electron-photon interactions yield effective binding of a recoiled electron pair via photon exchange, forming an electron pairing condensate with properties analogous to Cooper pairs in a lattice. The result indicates that the pairing instability eventually causes electron beams to bunch, a behavior similar to classical bunching in both accelerators and free-electron lasers. These findings suggest that light-modulated ultrafast electron beams can achieve a promising phase-coherent, superconducting-like phase of matter that suppresses unwanted space-charge repulsion, thereby offering a potential multi-particle ultrafast platform for quantum wavefunction engineering and opening new possibilities for free-electron quantum technologies.

**Beam bunching perspectives of free electron pairing.** Our result is schematically presented in Figure 1, which illustrates the critical steps for achieving light-induced pairing instability in ultrafast free electron beams, offering an intuitive framework for the discussions that follow. Panels 1a-1c presents the pairing process for plane-wave electrons. Our initial approach focuses on establishing a weak attractive interaction between PINEM electrons, inspired by the treatment in Ref. [28]. Following L. Cooper's seminal insight [27], any attractive forces between two free electrons guarantees the formation of a bound state, a phenomenon term "pairing instability" or the Cooper problem. In our setting, this attraction emerges through the Process of PINEM [11,12,32,33]. In this process, the free-electron wavefunction interacts with structured electromagnetic fields generated by surface plasmon polaritons (SPPs), which are driven by femtosecond laser pulses. This PINEM interaction facilitates the absorption and emission of multiple photons, potentially resulting in a net attractive force between electrons.

Typically, even when electron-photon coupling and second-order perturbation effects are considered, by tracing out the photon emission and absorption process in vacuum, the electron-electron interactions remain dominantly repulsive (Fig. 1b), resulting to the Coulomb forces derived from virtual photon fluctuation. However, when both PINEM electrons interact with structured photons mediated by SPP excitations (Fig. 1c), the laser-induced plasmonic dynamics modulate the quantum phase profile and sideband distribution of the incoming electrons, enabling a net attraction via the photon mediation. This phenomenon echoes Cooper's insight that when a net attraction exists between conduction band electrons, they can form bound states, and vice versa, this insight indicates a pathway toward achieving free-electron superconducting state for multi-particle PINEM electrons. In the following section, we will derive the effective attraction interaction using the Schrieffer–Wolff transformation [34], which can trace out all the participated photon degree of freedom to the second-order perturbation and obtain the photon-induced electron-electron interaction.

Our second perspective on free-electron pairing interprets the net attractive interaction from a classical acceleration viewpoint of electron beams, emphasizing the periodic microbunching of point-like electrons in presence of an electromagnetic field and illustrating how temporal bunching of paired electron wavefunctions resembles the pulse trains. In the bunching picture, the pairing behavior emerge from beam compression induced by the light field, with the net attraction results from periodic bunching of the beam. Within an optical cycle (see Fig. 1d), the phase-front electrons experience the negative electric field of the light to decelerate, therefore they emit photons. These emitted photons are then absorbed by the phase-behind electrons and being accelerated. To the second-order perturbation in QED framework, the connection between electrons by exchanging photons enable to produce an effectively quantum-mechanical re-bunching, which is to say, forming an attractive interaction among the micro-bunched electrons. At almost resonant and weak-field condition, this re-bunching induced attraction can mitigate the Coulomb repulsion between PINEM electrons, allowing pairing instability and further the condensation of the electrons in a bunch. As shown in Fig. 1e, we illustrate the processes of free-electron paring and paired electron condensation in the light-induced beam bunching process.

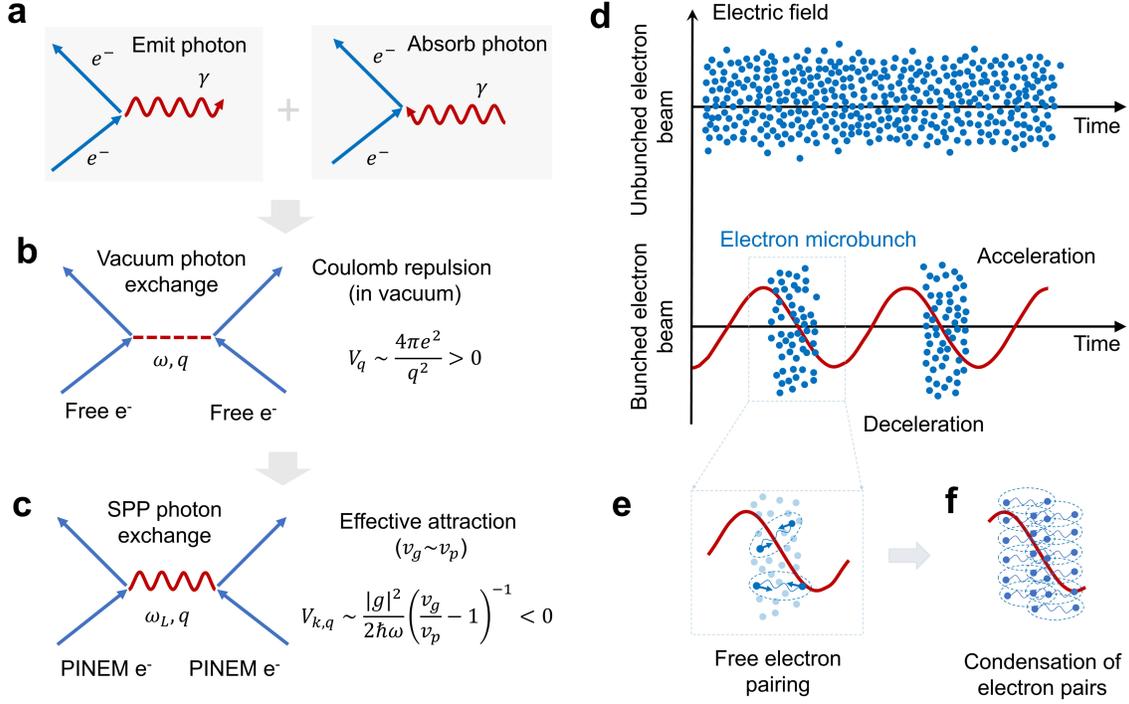

**Figure 1: Construction of light-induced free-electron pairing instability.** (a) shows the fundamental process of photon emission and absorption by electrons. (b) illustrates space-charge Coulomb repulsion via virtual photon exchange in QED framework. (c) The effective interaction between PINEM electrons by exchanging SPP photons. At the specific resonant regime, a net attraction between electrons can be obtained. Panels (d, e) present the pairing process in laser-induced particle acceleration process, with (d) comparing the longitudinal profiles of unbunched and bunched electron beams in presence of laser field. The demonstrates weak-field-limited photon exchanges between accelerated and decelerated electrons, leading to pairing instability during quantum bunching. (e) shows the quantum limit of laser acceleration, where photon exchange within microbunches enables electron pairing. Panels (d, e, f) depict the framework for establishing attractive interaction between PINEM electrons.

**Net Attraction using Schrieffer–Wolff transformation.** Our primary focus is on the PINEM interaction, a multiphoton process involving free electrons and optically near fields. Typically, in the PINEM process, an electron pulse containing roughly less than one electron is emitted to image the sample. However, in our study, we consider the multi-electron and light interaction, meaning that the electron pulses contain a large number of electrons that then couple to the structured electromagnetic field during the PINEM process. Specifically, the total Hamiltonian in multi-PINEM is expressed as [12,35],

$$\mathcal{H} = \mathcal{H}_0 + H_{ee} + H_{ep} \quad (1)$$

with the free part $\mathcal{H}_0 = H_e + H_p$, in which the free electron Hamiltonian is $H_e = \sum_k \varepsilon_k c_k^\dagger c_k$ and the free photon Hamiltonian is $H_p = \left(a_{q_L}^\dagger a_{q_L} + \frac{1}{2}\right)\hbar\omega_L$. Here, $c_k(c_k^\dagger)$ represents the free electron annihilation(creation) operator with momentum k in the $z$ direction, $a_{q_L}(a_{q_L}^\dagger)$ denotes the annihilation(creation) operator for near-field photons with frequency $\omega_L$ and wave number $q_L$. The electron energy is given by $\varepsilon_k = \varepsilon_0 + \hbar v_0(k-k_0) + \frac{\hbar^2(k-k_0)^2}{2\gamma^3 m}$, with $\varepsilon_0 = \gamma mc^2$, $\hbar k_0 = \gamma m v_0$, $v_0 = \beta c$ and the Lorentz factor $\gamma = \frac{1}{\sqrt{1-\beta^2}}$. The second term $H_{ee}$ represents the Coulomb interactions between electrons. Consider that within a uniformly distributed charge density traveling in free space at relativistic speed, both the transverse and longitudinal space charge forces vanish approximately as $1/\gamma^2$ due to the cancellation of the electric and magnetic forces and the Lorentz contraction along the z direction [36], we finally express the interactions between PINEM electrons as [see details in the Supplementary Materials]:

$$H_{ee} = \frac{1}{2}\sum_{k,k',n} V_{ee}^C(nq_L) c_{k+nq_L}^\dagger c_{k'-nq_L}^\dagger c_{k'} c_k \quad (2)$$

with $V_{ee}^C(nq_L) = \frac{Q}{\gamma^2 \epsilon_0} \frac{e^2}{(nq_L)^2 + \kappa^2}$, where $\epsilon_0$ is the vacuum permittivity and $\kappa$ denotes the momentum spread due to the transverse beam uncertainty, and $Q = 8\pi^3/\Omega$ is the normalized volume of the reciprocal space and $\Omega$ the volume of the real space. The last term $H_{ep}$ in Eq. (1) represents the electron-photon interactions as

$$H_{ep} = \sum_k \left(g a_{q_L} c_{k+q_L}^\dagger c_k + g^* a_{q_L}^\dagger c_k^\dagger c_{k+q_L}\right) = \sum_k (T_{k+} + T_{k-}) \quad (3)$$

where $T_{k+} = g a_{q_L} c_{k+q_L}^\dagger c_k$, $T_{k-} = T_{k+}^\dagger$, $g$ is the coupling factor between the photon and the electron, and the rotating wave approximation is applied with only the longitudinal electromagnetic mode considered [37]. Altogether, the total Hamiltonian is

$$\mathcal{H} = \mathcal{H}_0 + H_{ee} + H_{ep}$$
$$= \sum_k \varepsilon_k c_k^\dagger c_k + \left(a_{q_L}^\dagger a_{q_L} + \frac{1}{2}\right)\hbar\omega_L + \frac{1}{2}\sum_{k,k',n} V_{ee}^C c_{k+nq_L}^\dagger c_{k'-nq_L}^\dagger c_{k'} c_k + \sum_k \left(g c_{k+q_L}^\dagger c_k a_{q_L} + g^* c_k^\dagger c_{k+q_L} a_{q_L}^\dagger\right). \quad (4)$$

As demonstrated in Fig. 1c, we employ the Schrieffer–Wolff transformation (SWT) [38] to perform a second-order perturbation that integrate out the single-photon process, yielding the effective interaction between PINEM electrons. We apply the SWT to the total Hamiltonian to find an effective electron-photon decoupled Hamiltonian:

$$\begin{aligned}\widetilde{\mathcal{H}} =& \ e^S\mathcal{H}e^{-S} = \mathcal{H} + [S,\mathcal{H}] + \frac{1}{2!}[S,[S,\mathcal{H}]] + \cdots \\ =& \ \mathcal{H}_0 + H_{ee} + \left(H_{ep} + [S,\mathcal{H}_0]\right) + \frac{1}{2}\left[S, H_{ep} + [S,\mathcal{H}_0]\right] + \frac{1}{2}[S, H_{ep}] + \mathcal{O}(|g|^3)\end{aligned}$$

which is approximated as $\widetilde{\mathcal{H}} = \mathcal{H}_0 + H_{ee} + \frac{1}{2}[S, H_{ep}]$. Here, we choose the ansatz $S = \sum_k \frac{T_{k-} - T_{k+}}{\delta_k}$ with $\delta_k = \hbar\omega_L - (\varepsilon_{k+q_L} - \varepsilon_k)$, ensuring that $H_{ep} + [S,\mathcal{H}_0] = 0$. The standard procedure for the SWT is detailed in the SM file. To be specific, we assume the initial photon state is a Fock or coherent state with small photon number $\nu_0$. In the final step, we approximate that, under the quantum and weak field condition, the two-photon and higher-order processes can be neglected, such as $\langle \nu_0 | a_{q_L}^{(\dagger)} a_{q_L}^{(\dagger)} | \nu_0 \rangle \approx 0$, and this weak field approximation will make the pairing interactions dominate. By canceling the first order of electron-photon coupling, we obtain the effective second-order electron-electron interactions. Therefore, the final photon-mediated interaction Hamiltonian can be expressed as [see the SM file]

$$\begin{aligned}\widetilde{\mathcal{H}} =& \sum_k \varepsilon_k c_k^\dagger c_k + \frac{1}{2}\frac{Q}{\gamma^2 \epsilon_0} \sum_{k,k',n} \frac{e^2}{(nq_L)^2 + \kappa^2} c_{k+nq_L}^\dagger c_{k'-nq_L}^\dagger c_{k'} c_k \\ &- \frac{|g|^2}{2} \sum_{k,k'} \left(\frac{1}{\delta_k} + \frac{1}{\delta_{k'}}\right) c_{k'}^\dagger c_{k'+q_L} c_{k+q_L}^\dagger c_k \quad (5)\end{aligned}$$

where the effective Hamiltonian is now decoupled from the photon field now, and thus the free photon term is neglected in the following discussion. This SWT treatment minimizes the influence of high-energy processes, allowing for a clear analysis of the interaction between electrons, which shows that in PINEM setting, electrons are indirectly coupled through photon mediation, and this coupling manifests as effective interactions (Eq. 5) in low-energy scenarios.

In a driven many-body system the elementary absorption or emission of a photon with a wave vector $q_L$ and frequency $\omega_L$ is constrained by simultaneous conservation of energy and momentum. From these two conservation laws one obtains the resonant, or phase-matching, condition $\delta_k = 0$. In practice the external field can be detuned slightly from this exact resonance by varying either $\omega_L$ or the incidence angle that fixes $q_L$. When the detuning parameter $\delta_k \neq 0$, its sign dictates the character of the induced effective interaction. In the case $\delta_k < 0$ (electron energy transfer is smaller than the photon frequency), the structured optical field "keeps pace" with electron motion, inducing collective charge rearrangements that mediate attractive interactions (negative interaction amplitude in second-order perturbation theory), analogous to lattice distortions in phonon-mediated attraction. Conversely, for $\delta_k > 0$, the photon response lags, failing to screen Coulomb repulsion, leading to repulsive couplings—mirroring phonons when electrons outpace lattice oscillations. Hence by continuously tuning the phase-matching condition across $\delta_k = 0$, one can reversibly switch between attractive and repulsive interactions and, moreover, control their magnitude with high precision—an experimentally valuable knob for engineering many-body Hamiltonians under laser driving.

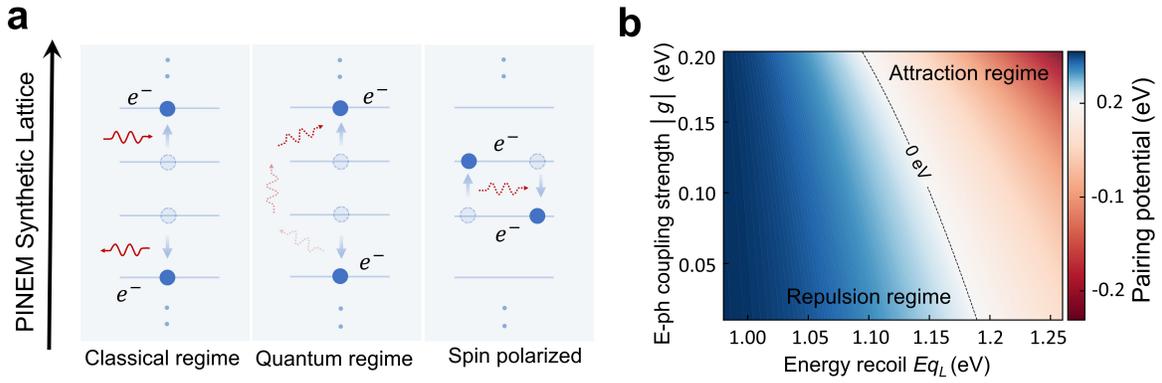

**Figure 2. Schematic illustration of the Hamiltonian describing photon-electron interactions.** (a) A single electron absorbs or emits a photon, leading to an increase or decrease in its momentum by $q_L$. This process can be viewed as a transition within the PINEM synthetic dimension. In the right, two electrons interact by absorbing and emitting the same photon, giving rise to an effective coupling between them. (b) When the electron group velocity and photon phase velocity experience specific detuning, the interplay between this pairwise interaction and the Coulomb interaction can lead to an attractive interaction in certain detuning ranges and coupling strength. Here we denote the recoil energy $E_{q_L} = \varepsilon_{k+q_L} - \varepsilon_k$, and take $Q = 0.0003 \text{ nm}^{-3}$ and $\hbar\omega_L = 1.4 eV$.

Let us take a close look at the detuning parameter $\delta_k$, which is explicitly expressed in the PINEM synthetic lattice (where $k = k_0 + nq_L$) as $\delta_k = \hbar v_0 q_L + (2n+1)\frac{\hbar^2 q_L^2}{2\gamma^3 m} - \hbar\omega_{q_L}$. And when the mismatch between the electron group velocity and the photon phase velocity is larger than the quadratic kinetic energy term of the electrons, i.e.,

$$v_p - v_0 \gg \frac{\hbar^2 q_L^2}{2\gamma^3 m},$$

where $v_p = \frac{\omega_{q_L}}{q_L}$ is the phase velocity of photons, and one sees that photon-mediated electron interaction becomes independent of momentum k and almost constant. With approximations, it leads to the effective interaction:

$$V_{eff} = \frac{|g|^2}{2}\sum_{k,k'} \frac{1}{\delta_k}\left(c_{k'}^\dagger c_{k'+q_L} c_{k+q_L}^\dagger c_k + c_k^\dagger c_{k+q_L} c_{k'+q_L}^\dagger c_{k'}\right) \simeq -\frac{V_0}{2}\sum_{k,k'} c_{k+q_L}^\dagger c_{k'-q_L}^\dagger c_{k'} c_k \quad (6)$$

where $V_0 = \frac{2|g|^2}{\hbar\omega_L - \hbar v_0 q_L} = \frac{2|g|^2}{\hbar q_L(v_p - v_0)}$. In the attractive regime $V_0 > 0$, where the photon phase velocity exceeds the electron group velocity, the "fast" photon acts as a mediator: state changes of one electron propagate via photon - mediated information transfer, inducing coordinated dynamical adjustments in the second electron. This synchronization arises because the photon outpaces electron motion, enabling interaction - driven state correlation through the field's ability to bridge electron dynamics at speeds exceeding the electron's group velocity. In the repulsive regime $V_0 < 0$, where the photon phase velocity lags behind the electron group velocity, the photon mediator fails to synchronize electron state dynamics: energy extraction from virtual transitions disrupts correlation, inducing repulsive interactions that prevent coordinated state adjustments between electrons.

To visualize the nature of pairing interactions for PINEM electrons, we express the many-body Hamiltonian in the synthetic PINEM basis, where the longitudinal momentum is quantized in units of photon momentum $\hbar q_L$. This construction gives rise to a "momentum lattice," as illustrated schematically in Fig. 2(a). In this lattice, each basis state with momentum $k + nq_L$ is connected to its nearest neighbors through single-photon absorption (n→n+1) or emission (n→n−1), resulting in a tight-binding kinetic term. In strong-field limit, photon absorption and emission events by individual electrons are independent and dominated by first-order photon-electron interactions, ultimately leading to the known PINEM process and electron density bunching, as shown in Fig. 1(d). In contrast, under weak-field regime, single electron and photon absorption and emission processes become suppressed and effective electron-electron pair interactions, as described by Eq. (6), become dominant. Here, two electrons can cooperatively absorb and emit the same (virtual) photon: for instance, if two electrons are initially separated in momentum space, one electron may emit a photon, approximately increasing its energy by $\hbar v_0 q_L$, and for a short interaction time, the other electron absorb this photon, transitioning to a higher energy sideband. This process leads to an attractive interaction between the electrons, as depicted in the central transition of Fig. 2(a). Furthermore, when the momentum difference between the two electrons is exactly $\hbar q_L$, as illustrated by the rightmost transition in Fig. 2(a), the pairing undergoes an exchange process, in which the two-electron wavefunction forms a bound state.

By adding the repulsive Coulomb term (keeping the first term $V(q_L)$ to the effective pairing interaction term, we obtain the total pairing potential

$$V_{\text{pairing}} = \frac{Q}{\gamma^2 \epsilon_0} \frac{e^2}{q_L^2 + \kappa^2} - \frac{|g|^2}{\hbar\omega_L - \hbar v_0 q_L} \quad (7),$$

which may become a net attractive potential when the parameters such as the electron-photon coupling strength $|g|$ and the recoil $E_{q_L}$ are tuned at some regimes, as shown in Fig. 2b.

**Formation of ultrafast free-electron bound states.** By summing over $k$ and $k'$ of Eq. (6), we obtain the Fourier-transformed Hamiltonian in real space:

$$V_{eff} = -\frac{V_0}{2} dz_1 dz_2 \psi^\dagger(z_1)\psi(z_1) \cos(q_L(z_1 - z_2)) \psi^\dagger(z_2)\psi(z_2). \quad (8)$$

From this result, we observe that the two-electron interaction is long-range, exhibiting periodic attractive and repulsive potentials. The Hamiltonian can be written explicitly as the sum of the free part and the interaction part $\widetilde{\mathcal{H}} = H_e + H_I$, with $H_I = \int dz_1 dz_2 \psi^\dagger(z_1)\psi(z_1)\psi^\dagger(z_2)\psi(z_2) V(z_1 - z_2)$ and $V(z) = -\frac{V_0}{2}\cos q_L z + \frac{e^2}{4\pi\epsilon_0\gamma^2} \frac{e^{-\kappa|z|}}{|z|}$. The real-space potential $V(z)$ reflects the fact that the photon-induced effective interaction can compensate for the space charge repulsion between electrons, as illustrated in Fig. 3(a).

To gain further physical intuition, we consider the case of two electrons. In the study of two-electron systems, the interaction Hamiltonian includes both the conventional Coulomb interaction and the photon-mediated interaction. For direct simulation, we define the corresponding Hamiltonian for two interacting electrons as follows:

$$H = H_1 + H_2 + V(z_1 - z_2) \quad (9)$$

where $H_{1,2}$ represents the kinetic energy of each electron, and $V(z_1 - z_2)$ is the potential energy due to interactions between the electrons. The kinetic energy terms are expressed as $H_{1,2} = E_0 + \frac{(p_{1,2} - p_0)^2}{2\gamma^3 m} + v_0(p_{1,2} - p_0)$ with $p_{1,2} = -i\hbar\partial_{z_{1,2}}$ being the momentum operators. The interaction potential is given by $V(z_1 - z_2) = \frac{e^2}{4\pi\varepsilon_0\gamma^2} \frac{e^{-\kappa|z_1-z_2|}}{|z_1-z_2|} - \frac{V_0}{2}\cos(q_L(z_1 - z_2))$, where the first term is the modified repulsive Coulomb potential and the second term is the photon-mediated electron interaction, with $V_0$ as the effective interaction strength. This photon-mediated interaction can effectively reduce the repulsive nature between electrons at specific distances, leading to possible bound states when the attractive interaction dominates.

Following our analysis of the two-electron Hamiltonian, we perform a transformation to center-of-mass and relative coordinates to simplify the time-dependent Schrodinger equation (TDSE) simulation. We define these coordinates as $Z = z_1 + z_2, \zeta = z_1 - z_2$, where $Z$ represents the center-of-mass coordinate and $z$ the relative coordinate. We rewrite the Hamiltonian (8) in terms of these new coordinates, leading to a separation of variables that simplifies the analysis. The transformed Hamiltonian is expressed as:

$$H = -i\hbar v_0\, \partial_Z - \frac{\hbar^2}{\gamma^3 m}\partial_Z^2 - \frac{\hbar^2}{\gamma^3 m}\partial_\zeta^2 + V(\zeta), \qquad (10)$$

where $V(\zeta)$ is the interaction potential dependent on the relative coordinate. This approach facilitates the study of the system by decoupling the center-of-mass motion from the relative motion of the particles.

Building on the two-electron real-space Hamiltonian in Eqs. (9) and (10), we investigate the dynamics of two electrons subject to an effective photon-mediated interaction $V(z_1 - z_2)$. As shown in Fig. 3a, the long-range potential—combined with the Coulomb interaction—yields an effective interaction profile that varies with position. We observe that for certain ranges of the photon momentum q and the position $\zeta$, two closely spaced electrons initially repel each other and then transition to an attractive regime. Once the separation exceeds λ (i.e., the relative distance $\zeta > \lambda$), the long-range interaction dominates, manifesting periodic intervals of repulsion and attraction regimes.

We further examine the evolution of the relative position of the two-electron wavepacket centers over time under different initial spatial separations. When the photon-induced effective interactions are neglected and only the space-charge interaction effects are considered, two electrons progressively separate due to mutual repulsive forces over time, as depicted in Fig. 3b. For cases with $V_0 > 0$, an appropriate coupling strength enables two nearby electron wavepackets to experience an attractive interaction. As a result, when two electrons are brought closer together by this attraction, the total energy of the system decreases, making it energetically favorable for them to form bound states. The alternating pattern of attraction and repulsion is illustrated in Figs. 3c, corresponding to strong photon-electron coupling with a deep attractive potential. Since $V(\zeta)$ represents a periodized long-range potential, if the decoherence length due to environmental noise in the PINEM system is much larger than both the electron wave packet length and the wavelength of the optical field, then attraction between distant electrons can also occur as can be seen in Fig. 3g. We see from Fig. 3g similar wavepackets evolution as Fig. 3c at a small range between in the relatively which leads to the possibility of long-range electron pairing in the free-electron system. Other cases with other spatial separations between the two electrons, such as the critical positions between the attractive and repulsive cases are shown in Figs. 3d-3f. The cases for attractive cases show specific small oscillations which mean that the two electron states are stable dynamically at the bottom of the attractive potentials.

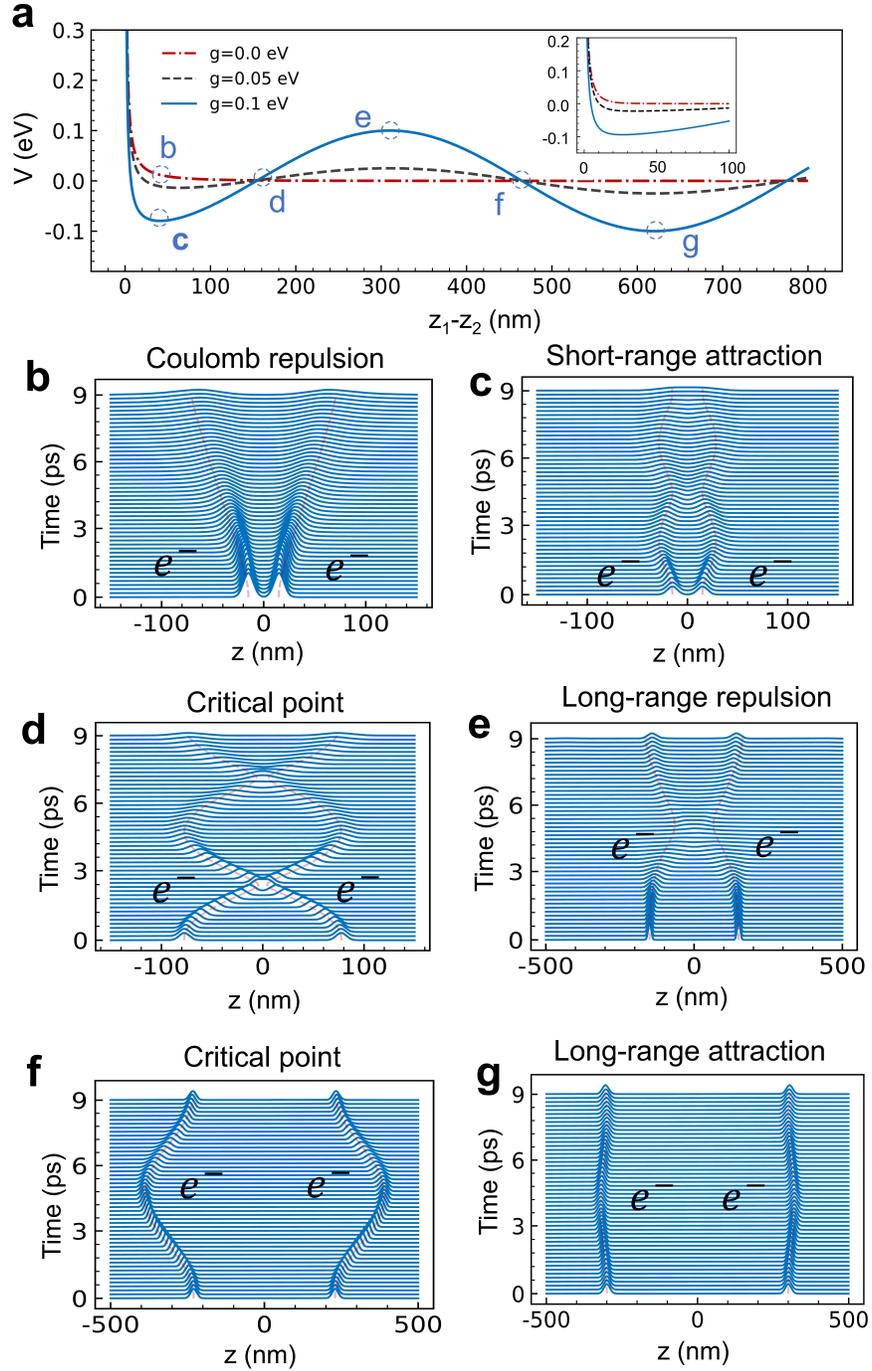

**Figure 3. Pairing interaction regimes and dynamical evolutions of two-electron bounded states.** (a) shows the total effective potential energy of the two-electron interaction under different coupling strengths as a function of relative position. When the spacing is small, the Coulomb interaction is dominant, resulting in a repulsive potential. As the spacing increases, the photon-mediated interaction becomes dominant, leading to an attractive potential. (b) When the coupling strength $|g| = 0$ and only space-charge repulsion potential remains. In (c-g), we take $|g| = 0.1 eV$ with different initial relative distances (as denoted in (a)) between two-electron wave packets. (c) and (g) show the electrons movement at the bottom of the negative potential well. (d) and (f) plot the electrons behavior at the critical point between negative and positive potential regions. In all cases, we take $\kappa = 0.1$ nm$^{-1}$.

**Promise from pairing instability to free-electron superconductor.** The formation of two-electron bound states provides insights into pairing and collective condensate phenomena. Transitioning from conventional superconductivity to free-electron superconductors via light-mediated pairing instability marks a fundamental shift in our understanding of quantum coherence among free electrons. In the context of photon-induced near-field electron microscopy (PINEM), electron beams are effectively one-dimensional and lack a well-defined "Fermi surface", complicating direct comparisons to traditional superconducting instability of Fermi liquids around Fermi surface [39]. It is more closely resemble to Frohlich's early model of one-dimensional superconductivity mediated by charge-density waves [40]. Nonetheless, by engineering strong coupling between ultrafast electron pulses and surface plasmon polariton (SPP) modes in the femtosecond regime, significant many-body interactions can be induced. These interactions may drive the electron beams into highly correlated bound states, as shown in Figs. 1e and 1f, where the emergence of free-electron pairs suggests the onset of a Bose-Einstein-condensate-like phase. This approach circumvents the limitations imposed by the absence of a Fermi surface, offering a novel route to many-body collective phenomena in ultrafast free-electron systems.

Notably, from an experimental standpoint, achieving and probing free-electron pairing and condensation in ultrafast bunched electron beams presents substantial challenges. The strong electron-photon coupling requires precise synchronization and advanced light pulse shaping [11,12]. Moreover, direct measurement of electron pair entanglement requires sophisticated and highly-sensitive detection schemes such as electron beam coincidence counting at sub-femtosecond time scale, tailored to periodically-bunched electron dynamics [30]. These technical advances are crucial for verifying pairing instability and for exploring emergent superconducting-like quantum coherence in free-electron platforms. Ultimately, overcoming these practical obstacles could enable a new platform that support free-electron pairing and collective behaviors analogous to superconductivity-within a fundamentally different, non-material system-leveraging the distinct advantages of free-electron beams for quantum information processing and ultrafast technologies.

**Conclusions**

In short, we have developed a theoretical framework demonstrating how ultrafast electron beams can be modulated by structured electromagnetic field to induce effective pairing instabilities, analogous to Cooper pairing in conventional superconductors. By leveraging near-resonant interactions between free electrons and structured optical fields, we reveal new mechanisms through which electron pairing and condensation may emerge in ultrafast electron beams. This opens promising directions for developing quantum coherent electron sources and for probing fundamental phenomena such as multi-electron correlations, quantum entanglement, and collective behavior in free-electron and light interactions. Taken together with our previous investigations on free electron topological phases [19], synthetic Bloch oscillations [35,41], the presented study supports the emergence of a new direction we refer to as free electron condensed matter physics. This vision aligns with the rapid progress in free-electron quantum optics [42] and suggests novel possibilities at the interaction of ultrafast optics, quantum information, and many-body physics.

**Acknowledgements.**

We thank Bin Zhang for helpful discussions. H.G. acknowledges the support of the NSFC (No. 12304068) and the startup Fund of Nanjing University of Aeronautics and Astronautics Grant No. YAH24076. J. K. acknowledges the support from the NSFC Grant No. 12074276, the Double First-Class Initiative Fund of ShanghaiTech University, and the start-up grant of ShanghaiTech University. Y.P. acknowledges the support of the NSFC (No. 2023X0201-417-03) and the fund of the ShanghaiTech University (Start-up funding).

The authors declare no competing financial interests.

# Supplementary Material for

# Light-induced Pairing Instability of Ultrafast Free Electrons with Space Charge Interactions

Hao Geng[1,2], Qiaofei Pan[3,4], Jian Kang[5], Yiming Pan[5,6*]

1. College of Physics, Nanjing University of Aeronautics and Astronautics, Nanjing 211106, China

2. Key Laboratory of Aerospace Information Materials and Physics (NUAA), MIIT, Nanjing 211106, China

3. Institute of Precision Optical Engineering, School of Physics Science and Engineering, Tongji University, Shanghai, 200092, China

4. MOE Key Laboratory of advanced micro-structure materials, Shanghai, 200092, China

5. State Key Laboratory of Quantum Functional Materials, School of Physical Science and Technology, ShanghaiTech University, Shanghai 200031, China

6. Center for Transformative Science, ShanghaiTech University, Shanghai 200031, China

**This pdf file includes two sections:**
(1) Schrieffer-Wolff Transformation for Electron-Photon Interaction.
(2) Two-Electron Schrödinger Equation in Real Space

## Schrieffer-Wolff Transformation for Electron-Photon Interaction

In this section, we derive the Schrieffer-Wolff transformation (SWT) [1] applied to the electron-photon interaction Hamiltonian. The goal is to systematically eliminate the interaction term to obtain an effective electron-electron interaction mediated by photons.

We define the operators and symbols used throughout the derivation: $c_k$ and $c_k^\dagger$ denote the fermionic annihilation and creation operators with momentum $k$, respectively; $a_{q_L}$ and $a_{q_L}^\dagger$ denote the bosonic annihilation and creation operators with momentum $q_L$; $g$ is the complex coupling constant; and $e_z$ is the unit vector along the $z$-direction. These operators satisfy the following commutation and anticommutation relations: the fermionic anticommutation relations

$$\{c_k, c_k'^\dagger\} = \delta_{k,k'}, \quad \{c_k, c_k'\} = 0, \quad \{c_k^\dagger, c_k'^\dagger\} = 0,$$

and the bosonic commutation relations

$$[a_{q_L}, a_{q_L}^\dagger] = 1, \quad [a_{q_L}, a_{q_L}] = 0, \quad [a_{q_L}^\dagger, a_{q_L}^\dagger] = 0.$$

Moreover, we recall some basic operator identities: for any operators $A$, $B$, and $C$, the commutators satisfy

$$[AB, C] = A[B, C] + [A, C]B, \quad \text{and} \quad [A, BC] = [A, B]C + B[A, C],$$

while for fermionic operators the anticommutator relation

$$\{AB, C\} = A\{B, C\} - \{A, C\}B.$$

We begin with the total Hamiltonian

$$\mathcal{H} = \mathcal{H}_0 + H_{ee} + H_{ep},$$

where the free (non-interacting) part is defined as

$$\mathcal{H}_0 = H_e + H_p.$$

Below we explain in detail the meaning of each parameter appearing in the following formulas.

**Free-Electron Hamiltonian:** The free-electron Hamiltonian, describing the one-dimensional relativistic dynamics along the $z$-direction (with spin neglected), is given by

$$H_e = \sum_k \varepsilon_k\, c_k^\dagger c_k,$$

where $k$ is the wavevector along the $z$-direction, and $c_k$ ($c_k^\dagger$) denotes the fermionic annihilation (creation) operator. The energy dispersion relation is

$$\varepsilon_k = \varepsilon_0 + \hbar v_0 (k - k_0) + \frac{\hbar^2 (k - k_0)^2}{2\gamma^3 m},$$

with the parameters defined as follows: the rest energy offset $\varepsilon_0 = (\gamma - 1)mc^2$; the initial electron velocity $v_0 = \beta c$, with $\beta$ being the ratio of the electron speed to the speed of light and $\gamma = \frac{1}{\sqrt{1-\beta^2}}$ the Lorentz factor; $k_0$ is a reference wavevector, and $m$ is the electron mass (with $c$ as the speed of light and $\hbar$ the reduced Planck constant). In typical PINEM setups, these are chosen as $\varepsilon_0 = 200\,keV$, $\beta = 0.7$, and $\gamma = 1.4$. In addition, the transverse kinetic energy

$$\frac{\hbar^2 (k_x^2 + k_y^2)}{2\gamma^3 m}$$

is neglected as its contribution is negligible compared to the longitudinal part (or it can be absorbed into the $\varepsilon_0$).

**Electron-Electron (Space charge) Interaction:** The space charge interaction between electrons in real space is modeled by

$$H_{ee} = \frac{1}{2\int d\mathbf{r}} d\mathbf{r}'\, V_e(\mathbf{r} - \mathbf{r}') \psi^\dagger(\mathbf{r}) \psi(\mathbf{r}) \psi^\dagger(\mathbf{r}') \psi(\mathbf{r}'),$$

with the Coulomb potential given by

$$V_e(\mathbf{r}) = \frac{1}{4\pi\gamma^2 \epsilon_0} \frac{e^2}{|\mathbf{r}|}.$$

Here $\epsilon_0$ represents the vacuum permittivity (e.g., $8.854 \times 10^{-12}\,\frac{F}{m}$ or $55.263\,e^2\,eV^{-1}\mu m^{-1}$), and $e$ is the elementary charge.

In momentum space the Coulomb interaction takes the form

$$H_{ee} = \frac{1}{2} \sum_{k,k',q} V_e(\mathbf{q})\, c_{k+q}^\dagger c_{k'-q}^\dagger c_{k'} c_k,$$

and when focusing on the dominant longitudinal modes, it is simplified to

$$H_{ee} = \frac{1}{2} \sum_{k,k',n} V'_e(nq_L) c^\dagger_{k+nq_L} c^\dagger_{k'-nq_L} c_{k'} c_k,$$

where $q_L$ is the longitudinal momentum transfer and

$$V'_e(nq_L) = \frac{Q}{\gamma^2 \epsilon_0} \frac{e^2}{(nq_L)^2 + \kappa^2}$$

with $\kappa^2$ representing the momentum spread due to transverse beam uncertainty and $Q = \frac{8\pi^3}{\Omega}$ being the normalized volume in the reciprocal space where $\Omega$ is the normalized volume in real space.

**Photon Hamiltonian:** The free photon Hamiltonian is defined as

$$H_p = \left(a^\dagger_{q_L} a_{q_L} + \frac{1}{2}\right) \hbar \omega_L$$

where $a_{q_L}$ ($a^\dagger_{q_L}$) is the bosonic annihilation (creation) operator for photons with momentum $q_L$ along the $z$-direction, and $\omega_L$ is the light frequency.

**Electron-Photon Coupling:** In our treatment of light–matter interaction, the electron–photon coupling Hamiltonian is written as

$$H_{ep} = -\int d\mathbf{r} \frac{e}{2\gamma m} \psi^\dagger(\mathbf{r})(p \cdot A + A \cdot P)\psi(\mathbf{r})$$

where $\psi(\mathbf{r})$ is the electron field operator, $p$ denotes the momentum operator, $A$ is the vector potential describing the quantized electromagnetic field, $e$ is the electron charge, $m$ is the electron mass, and $\gamma$ is the Lorentz factor. For a structured optical field—as used in PINEM experiments where the longitudinal field component is predominant—the vector potential is expanded in terms of free-space plane-wave modes:

$$A(\mathbf{r},t) = \sum_{q,\lambda} \sqrt{\frac{\hbar}{2\epsilon_0 \omega_q \mathcal{V}}} \left(a_{q\lambda} e^{i(qz-\omega_q t)} + a^\dagger_{q\lambda} e^{-i(qz-\omega_q t)}\right) \mathbf{e}_\lambda$$

with $a_{q\lambda}$ and $a^\dagger_{q\lambda}$ being the photon annihilation and creation operators for the mode characterized by momentum $q$ and polarization $\lambda$, $\mathbf{e}_\lambda$ the corresponding polarization unit vector, $\omega_q$ the angular frequency, and $\mathcal{V}$ the effective volume of the electric field. In this expansion the eigenmode amplitude is identified as

$$\tilde{E}_{q\lambda} = \sqrt{\frac{\hbar\omega_q}{2\epsilon_0 \mathcal{V}}},$$

which sets the overall scale of the interaction via the coupling factor which comes from the quantum vacuum zero-point energy.

Within the rotating-wave approximation—valid when $|g| \ll \hbar v_0 q_L \approx \hbar\omega_L$—this treatment leads to a simplified electron–photon coupling Hamiltonian,

$$\mathrm{H}_{ep} = \sum_k g a_{q_L} c^\dagger_{k+q_L} c_k + g^* a^\dagger_{q_L} c^\dagger_k c_{k+q_L}$$

with the coupling factor defined as

$$g = -\frac{e\tilde{E}_{q_L}}{2\gamma m \omega_L} p_0 e^{i\phi} = -\frac{e}{2\gamma m}\sqrt{\frac{\hbar}{2\omega_L \epsilon_0 \mathcal{V}}} p_0 e^{i\phi},$$

where $p_0$ is the electron momentum and $\phi$ is the phase of the mode. This formulation consistently incorporates both the correct field normalization in vacuum and the electron dynamics relevant for structured light–matter interactions.

**The Schrieffer–Wolff Transformation:** We aim to eliminate $H_{ep}$ using a unitary transformation. For the unitary operator $\mathcal{U}_S = e^S$ (where $S$ is anti-Hermitian), the transformed Hamiltonian is expressed as:

$$\tilde{\mathcal{H}} = e^S \mathcal{H} e^{-S} = \mathcal{H} + [S, \mathcal{H}] + \frac{1}{2}[S, [S, \mathcal{H}]] + \cdots.$$

The interaction term $H\_ep$ can be decomposed as:

$$H_{ep} = \sum_k (T_{k+} + T_{k-}),$$

where

$$T_{k+} = g a_{q_L} c^\dagger_{k+q_L} c_k, \quad T_{k-} = g^* a^\dagger_{q_L} c^\dagger_k c_{k+q_L}, \quad T^\dagger_{k+} = T_{k-}.$$

To cancel $H_{ep}$ at the leading order, we propose $S$ as a linear combination of $T_{k+}$ and $T_{k-}$:

$$S = \sum_k (f_k T_{k+} + h_k T_{k-}).$$

First, we calculate the commutator $[T_{k+}, H_e]$:

$$[T_{k+}, H_e] = \sum_{k'} [g a_{q_L} c^\dagger_{k+q_L} c_k, \varepsilon_{k'} c^\dagger_{k'} c_{k'}]$$

$$= g a_{q_L} \sum_{k'} \varepsilon_{k'} [c^\dagger_{k+q_L} c_k, c^\dagger_{k'} c_{k'}].$$

Using the commutation relation:

$$[c^\dagger_{k+q_L} c_k, c^\dagger_{k'} c_{k'}] = c^\dagger_{k+q_L} \{c_k, c^\dagger_{k'} c_{k'}\} - \{c^\dagger_{k+q_L}, c^\dagger_{k'} c_{k'}\} c_k,$$

we find:

$$[c^\dagger_{k+q_L} c_k, c^\dagger_{k'} c_{k'}] = \delta_{k',k} c^\dagger_{k+q_L} c_{k'} - \delta_{k+q_L,k'} c^\dagger_{k'} c_k.$$

Substituting this result, we obtain:

$$[T_{k+}, H_e] = g a_{q_L} (\varepsilon_k - \varepsilon_{k+q_L}) c^\dagger_{k+q_L} c_k.$$

Next, we compute the commutator $[T_{k+}, H_p]$:

$$[T_{k+}, H_p] = g[a_{q_L} c^\dagger_{k+q_L} c_k, \hbar \omega_L a^\dagger_{q_L} a_{q_L}]$$

$$= g \hbar \omega_L [a_{q_L}, a^\dagger_{q_L} a_{q_L}] c^\dagger_{k+q_L} c_k.$$

Using the commutation relation $[a_{q_L}, a^\dagger_{q_L} a_{q_L}] = a_{q_L}$, we find:

$$[T_{k+}, H_p] = g a_{q_L} \hbar \omega_L c^\dagger_{k+q_L} c_k.$$

Combining the results for $H\_e$ and $H\_p$, we have:

$$[T_{k+}, \mathcal{H}_0] = g a_{q_L} (\hbar \omega_L + \varepsilon_k - \varepsilon_{k+q_L}) c^\dagger_{k+q_L} c_k.$$

If we choose

$$f_k = \frac{1}{\varepsilon_{k+q_L} - \varepsilon_k - \hbar \omega_L},$$

then

$$[f_k T_{k+}, \mathcal{H}_0] = -T_{k+}.$$

Now, we calculate $[T_{k-}, \mathcal{H}_0]$, which is given by the sum of $[T_{k-}, H_e]$ and $[T_{k-}, H_p]$. First, consider $[T_{k-}, H_e]$:

$$[T_{k-}, H_e] = \sum_{k'} [g^* a_{q_L}^\dagger c_k^\dagger c_{k+q_L}, \varepsilon_{k'} c_{k'}^\dagger c_{k'}]$$

$$= g^* a_{q_L}^\dagger \sum_{k'} \varepsilon_{k'} [c_k^\dagger c_{k+q_L}, c_{k'}^\dagger c_{k'}].$$

Using the commutation relation:

$$[c_k^\dagger c_{k+q_L}, c_{k'}^\dagger c_{k'}] = c_k^\dagger c_{k'} \delta_{k+q_L, k'} - c_{k'}^\dagger c_{k+q_L} \delta_{k, k'},$$

we find:

$$[T_{k-}, H_e] = g^* a_{q_L}^\dagger (\varepsilon_{k+q_L} - \varepsilon_k) c_k^\dagger c_{k+q_L}.$$

Next, compute $[T_{k-}, H_p]$:

$$[T_{k-}, H_p] = -g^* \hbar \omega_L a_{q_L}^\dagger c_k^\dagger c_{k+q_L}.$$

Combining these results, we have:

$$[T_{k-}, \mathcal{H}_0] = (\varepsilon_{k+q_L} - \varepsilon_k - \hbar \omega_L) g^* a_{q_L}^\dagger c_k^\dagger c_{k+q_L}.$$

If we choose

$$h_k = -\frac{1}{\varepsilon_{k+q_L} - \varepsilon_k - \hbar \omega_L},$$

then

$$h_k [T_{k-}, \mathcal{H}_0] = -T_{k-}.$$

Since $H_{ep} = \sum_k (T_{k+} + T_{k-})$, we find:

$$H_{ep} + [S, \mathcal{H}_0] = 0.$$

Thus, $S$ can be expressed as:

$$S = \sum_k \frac{T_{k-} - T_{k+}}{\delta_k},$$

where

$$\delta_k = -\frac{1}{f_k} = \frac{1}{h_k} = \hbar\omega_L - (\varepsilon_{k+q_L} - \varepsilon_k).$$

Using the first-order expansion of the transformation, $H\_ep$ is eliminated, and we write:

$$\tilde{\mathcal{H}}_0 + \tilde{H}_{ep} = \mathcal{H}_0 + \frac{1}{2}[S, H_{ep}] + \mathcal{O}(|g|^3).$$

To compute $[S, H\_ep]$, we evaluate:

$$[S, H_{ep}] \approx -\sum_{k,k'} \frac{1}{\delta_k} |g|^2 \left( c_{k'}^\dagger c_{k'+q_L} c_{k+q_L}^\dagger c_k + c_k^\dagger c_{k+q_L} c_{k'+q_L}^\dagger c_{k'} \right).$$

Here we assume the system is in the weak light regime, where the photon number $\nu_0$ is small, and the $< a_{q_L}^{(\dagger)} (a_{q_L}^{(\dagger)}) > \mu_0$ can be neglected. The effective electron-electron interaction mediated by photons is given by:

$$H_{eff} = \frac{1}{2}[S, H_{ep}].$$

Finally, the transformed Hamiltonian becomes:

$$\tilde{\mathcal{H}} \approx \mathcal{H}_0 + H_{ee} + H_{eff},$$

where $H\_ee$ is the space charge interaction in momentum space:

$$H_{ee} = \frac{1}{2} \sum_{k,k',n} V_e'(nq_L)\, c_{k+nq_L}^\dagger\, c_{k'-nq_L}^\dagger\, c_{k'}\, c_k,$$

with $V_e'(nq_L) = \frac{Q}{\gamma^2 \epsilon_0} \frac{e^2}{(nq_L)^2 + \kappa^2}$, and $H_{eff}$ captures the effective interaction induced by the photon field. The effective interaction can be expressed as:

$$H_{eff} = -\frac{|g|^2}{2} \sum_{k,k'} \frac{1}{\delta_k} (c_{k'}^\dagger\, c_{k'+q_L}\, c_{k+q_L}^\dagger\, c_k + c_k^\dagger\, c_{k+q_L}\, c_{k'+q_L}^\dagger\, c_{k'}).$$

For the specific case where $n = 1$ in the space charge interaction, $H\_ee$ becomes:

$$H_{ee}^{(n=1)} = \frac{1}{2} \sum_{k,k'} V_e'(q_L) c_{k+q_L}^\dagger c_{k'-q_L}^\dagger c_{k'} c_k,$$

with $V_e'(q_L) = \frac{Q}{\gamma^2 \epsilon_0} \frac{e^2}{q_L^2 + \kappa^2}$.

We can now combine this $n = 1$ space charge interaction with the photon-mediated effective interaction to obtain a total pairing interaction:

$$H_{pair} = H_{ee}^{(n=1)} + H_{eff} = \frac{1}{2} \sum_{k,k'} V_{pair}(k,k') c_{k+q_L}^\dagger c_{k'-q_L}^\dagger c_{k'} c_k,$$

where the effective pairing potential is given by:

$$V_{pair}(k,k') = \frac{Q}{\gamma^2 \epsilon_0} \frac{e^2}{q_L^2 + \kappa^2} - \frac{|g|^2}{\hbar \omega_L - (\varepsilon_{k+q_L} - \varepsilon_k)}.$$

This combined interaction represents the total effective electron-electron interaction that includes both the direct Coulomb repulsion and the photon-mediated attraction. The competition between these two mechanisms determines whether the overall interaction is attractive or repulsive, which is crucial for understanding the pairing instability in the system.

In the ultrafast limit, the square term of kinetic energy can be neglected, and $\varepsilon_{k+q_L} - \varepsilon_k = \hbar v_0 q_L$, which corresponds to the phase match condition in PINEM experiments, the pairing potential becomes:

$$V_{pair}(k,k') = \frac{Q}{\gamma^2 \epsilon_0} \frac{e^2}{q_L^2 + \kappa^2} - \frac{|g|^2}{\hbar \omega_L - \hbar v_0 q_L}.$$

When the laser frequency is tuned such that $\omega_L = v_0 q_L$, the denominator of the second term approaches zero, leading to a significant enhancement of the photon-mediated interaction. In this case, the effective pairing potential can be approximated as:

$$V_{pair}(k,k') \approx \frac{Q}{\gamma^2 \epsilon_0} \frac{e^2}{q_L^2 + \kappa^2} - \frac{|g|^2}{\delta_0},$$

where $\delta_0 = \hbar q_L \left(\frac{\omega_L}{q_L} - v_0\right)$ is a small detuning parameter. This expression clearly shows how the photon-mediated attraction (negative term) can overcome the Coulomb repulsion (positive term) when the system is near resonance, potentially leading to electron pairing.

**Two-Electron Schrödinger Equation in Real Space**

The two-particle system is described by the time-dependent Schrödinger equation in the first quantization as:

$$i\hbar\, \partial_t \psi(z_1, z_2, t) = H\psi(z_1, z_2, t),$$

where the Hamiltonian is given by:

$$H = \hat{H}_1 + \hat{H}_2 + \hat{V}(z_1 - z_2),$$

with

$$\hat{H}_{1,2} = E_0 + \frac{(\hat{p}_{1,2} - p_0)^2}{2\gamma^3 m} + v_0(\hat{p}_{1,2} - p_0), \quad \hat{p}_{1,2} = -i\hbar\, \partial_{z_{1,2}}.$$

The commutation relation is:

$$[z_i, \hat{p}_j] = i\hbar \delta_{i,j}.$$

Here, $\hat{V}(z_1 - z_2)$ represents the interaction potential, which may include contributions from electron-photon coupling.

To simplify the dynamics, we apply a unitary transformation:

$$\tilde{\psi} = (z_1, z_2, t) = U_E(t)\psi(z_1, z_2, t),$$

where

$$U_E(t) = e^{\frac{i}{\hbar}(2E_0 t - p_0 z_1 - p_0 z_2)}.$$

The transformed Hamiltonian becomes:

$$\tilde{H} = U_E(t) H U_E^\dagger(t) + i\hbar\big(\partial_t U_E(t)\big) U_E^\dagger(t),$$

which simplifies to:

$$\tilde{H} = \sum_{i=1,2}\left(v_0\hat{p}_i + \frac{\hat{p}_i^2}{2\gamma^3 m}\right) + V(z_1 - z_2),$$

where $V(z_1 - z_2)$ is the interaction potential which include contributions from electron-photon coupling and Coulomb interaction and it is given by:

$$V(z_1 - z_2) = \frac{1}{4\pi\gamma^2\epsilon_0}\frac{e^2 e^{-\kappa|z_1-z_2|}}{|z_1 - z_2|} - \frac{|g|^2}{\delta_0}\cos(z_1 - z_2)$$

The corresponding Schrödinger equation is:

$$i\hbar\,\partial_t \tilde{\psi}(z_1, z_2, t) = \tilde{H}\tilde{\psi}(z_1, z_2, t).$$

To further simplify, we introduce center-of-mass $Z$ and relative $\zeta$ coordinates:

$$Z = \frac{z_1 + z_2}{2}, \quad \zeta = z_1 - z_2,$$

with the Jacobian matrix $A$ and its inverse transpose $B$:

$$A = \begin{pmatrix} \frac{1}{2} & \frac{1}{2} \\ 1 & -1 \end{pmatrix}, \quad B^T = A^{-1} = \begin{pmatrix} 1 & 1 \\ \frac{1}{2} & -\frac{1}{2} \end{pmatrix}.$$

The derivatives transform as:

$$\begin{pmatrix} \partial_{z_1} \\ \partial_{z_2} \end{pmatrix} = A^T \begin{pmatrix} \partial_Z \\ \partial_\zeta \end{pmatrix}.$$

Substituting these coordinates into $\tilde{H}$, we find:

$$\tilde{H} = -i\hbar v_0\,\partial_Z - \frac{\hbar^2}{4\gamma^3 m}\partial_Z^2 - \frac{\hbar^2}{\gamma^3 m}\partial_\zeta^2 + V(\zeta).$$

By applying the transformation $Z' = Z - v_0 t$ and defining:

$$\Phi(Z', \zeta, t) = \Psi(Z + v_0 t, \zeta, t),$$

the equation becomes:

$$i\hbar\,\partial_t \Phi = \left(-\frac{\hbar^2}{4\gamma^3 m}\partial_Z^2 - \frac{\hbar^2}{\gamma^3 m}\partial_\zeta^2 + V(\zeta)\right)\Phi.$$

where we use $Z$ to denote the comoving center of mass coordinate which is just $Z'$ in the above.

At $t = 0$, the initial Gaussian wavepacket in terms of $z_1, z_2$ coordinates is:

$$\Phi(Z, \zeta, 0) = \frac{1}{\Omega} e^{-\frac{\left(Z+\frac{\zeta}{2}-Z_{01}\right)^2}{\sigma^2} - \frac{\left(Z-\frac{\zeta}{2}-Z_{02}\right)^2}{\sigma^2} + i\left(k_1\left(Z+\frac{\zeta}{2}\right) + k_2\left(Z-\frac{\zeta}{2}\right)\right)}.$$

Choosing $Z_{01} = -Z_{02} = \frac{Z_0}{2}$ (separation $Z_0$), the wavepacket separates into:

$$\Phi(Z, \zeta, 0) = \phi_1(Z, 0)\phi_2(\zeta, 0),$$

where

$$\phi_1(Z, 0) = \frac{1}{\Omega} e^{-\frac{2Z^2}{\sigma^2} + i(k_1+k_2)Z},$$

and

$$\phi_2(\zeta, 0) = e^{-\frac{(\zeta-Z_0)^2}{2\sigma^2} + i\frac{(k_1-k_2)}{2}\zeta}.$$

The Schrödinger equation now separates into two independent equations:

$$i\hbar\, \partial_t \phi_1 = -\frac{\hbar^2}{4\gamma^3 m} \partial_Z^2 \phi_1,$$

$$i\hbar\, \partial_t \phi_2 = \left(-\frac{\hbar^2}{\gamma^3 m} \partial_\zeta^2 + V(\zeta)\right) \phi_2,$$

with initial conditions:

$$\phi_1(Z, 0) = \frac{1}{\Omega} e^{-\frac{2Z^2}{\sigma^2} + i(k_1+k_2)Z},$$

$$\phi_2(\zeta, 0) = e^{-\frac{(\zeta-Z_0)^2}{2\sigma^2} + i\frac{(k_1-k_2)}{2}\zeta}.$$

For the center-of-mass motion $\phi_1(Z, t)$, the solution is a spreading Gaussian wavepacket:

$$\phi_1(Z, t) = \frac{1}{\Omega} e^{-2\frac{(Z-v_g t)^2}{\sigma^2\left(1+\frac{i\hbar t}{\gamma^3 m \sigma^2}\right)} + i(k_1+k_2)\left(Z - \frac{\hbar(k_1+k_2)t}{4\gamma^3 m}\right)},$$

where $v_g = \frac{\hbar(k_1+k_2)}{2m\gamma^3}$ is the group velocity of the electron.